# Reconnecting Fragmented Citation Networks with Semantic Augmentation

*Vu Thi Huong*[*12], *Annika Buchholz*[1], *Imene Khebouri*[1], *Thorsten Koch*[34], *Tim Kunt*[1], *Wolfgang Peters-Kottig*[5], *Tomasz Stompor*[5], *Janina Zittel*[4]

Emails: {huong.vu, buchholz, khebouri, koch, kunt, peters-kottig, stompor, zittel}@zib.de

[1] Digital Data and Information for Society, Science, and Culture,
Zuse Institute Berlin, Takustr. 7, 14195 Berlin, Germany

[2] Institute of Mathematics,
Vietnam Academy of Science and Technology, 10072 Hanoi, Vietnam

[3] Software and Algorithms for Discrete Optimization,
Technische Universität Berlin, Straße des 17. Juni 135, 10623 Berlin, Germany

[4] Applied Optimization,
Zuse Institute Berlin, Takustr. 7, 14195 Berlin, Germany

[5] Kooperativer Bibliotheksverbund Berlin-Brandenburg (KOBV),
Zuse Institute Berlin, Takustr. 7, 14195 Berlin, Germany

[*] *Corresponding author*

**Abstract.** Citation graphs are fundamental tools for modeling scientific structure, but are often fragmented due to missing citations of scientifically connected articles. To address this issue, we propose a computationally efficient hybrid framework integrating citation topology with large language model (LLM)-based text similarity. Using 662,369 Web of Science publications in Mathematics and Operations Research & Management Science, we augment the original graph by adding semantic edges from small, disconnected components and weighting existing citations according to textual similarity. Semantic augmentation substantially reduces fragmentation while preserving disciplinary homogeneity. Compared to embedding-only clustering, cluster detection on augmented graphs using the Leiden algorithm retains structural interpretability while offering multi-scale organization. The method scales efficiently to large datasets and offers a practical strategy for strengthening citation-based indicators without collapsing disciplinary boundaries.

## 1. Introduction

Citation networks form the backbone of quantitative science mapping. They are used to delineate fields, normalize impact indicators, detect emerging topics, and analyze interdisciplinarity. Community detection in citation graphs has become a standard tool for identifying scientific structures at multiple scales; see (Šubelj, et al., 2016; Waltman, et al., 2020) and the references therein.

Citations themselves have a strong positive epistemic function. They are not randomly distributed links but selective acknowledgments made by domain experts. Because researchers cite prior work they consider relevant, foundational, or methodologically important, citation links often reflect informed judgments about intellectual influence and topical proximity. In this sense, citations provide a structured and expert-curated signal that helps organize the scientific knowledge space. While individual citations may be imperfect, in aggregate they offer a meaningful indicator for structuring scientific networks and tracing the evolution of research areas.

However, citation graphs are not complete representations of intellectual structure. They could also be shaped by several constraints, including:

- database coverage limitations,
- missing or unmatched references,
- time-window truncation,
- extraction artifacts.

These factors imply that citation-based networks approximate, rather than fully capture the underlying organization of science.

As a result, observed citation networks are structurally fragmented. In large-scale bibliometric datasets, this fragmentation often yields small, disconnected components that may not correspond to meaningful epistemic communities. Community detection algorithms, such as the Leiden algorithm (Traag, et al., 2019), interpret structural disconnection as a substantive separation. This poses a methodological concern about academic indicators. Field delineation, normalization schemes, and measures of interdisciplinarity depend on stable and structurally coherent partitions. If network fragmentation is partly an artifact of incomplete linkage, downstream indicators may inherit structural distortions.

In addition to citation information, titles and abstracts are also carefully crafted by authors to summarize a paper's key contributions, context, and scope, making them highly informative sources of topical signals. Recent advances in large language models (LLMs) enable us to extract dense semantic representations (Petukhova, et al., 2025) from these summaries, thereby capturing distilled expertise and enabling meaningful comparisons between papers even in the absence of citation links. Unlike citation graphs, semantic similarity spaces are dense by construction, so any two documents can be compared. This raises a key methodological question:

> ***Can semantic similarity compensate for missing citation edges while preserving the interpretability and disciplinary structure of citation networks?***

Rather than replacing citation-based clustering with semantic-embedding-only approaches, we propose a **scalable hybrid strategy** in this work. Citation topology remains the structural backbone of the network while semantic similarity is augmented to repair likely structural fragmentation and refine edge strength.

Using 662,369 Web of Science (WoS) publications, we evaluate whether semantic augmentation:

- reduces artificial fragmentation,
- preserves disciplinary homogeneity ,
- remains computationally feasible at scale.

Our results show that targeted semantic augmentation substantially reduces small disconnected components while maintaining high subject homogeneity. Compared to embedding-only clustering, the hybrid approach preserves structural interpretability and multi-scale organization.

## 2. Data and Baseline Structure

The dataset comprises 662,369 publications indexed in WoS (2000-2024) in two subject categories: *Mathematics* and *Operations Research & Management Science* (OR&MS). It builds on the versions used in (Huong, et al., 2026) and (Huong & Koch, 2025), in which non-English publications, records missing essential metadata (e.g., WoS unique identifier, title, or publication year), and entries with inconsistent self-citations were removed. After these steps, the largest connected component was retained and nodes with degree one were excluded. In this work, we further refined the dataset by removing publications whose abstracts contain fewer than 100 characters to improve the reliability of textual similarity extraction. The final citation network consists of 4,150,852 directed edges representing citing-cited publication pairs.

For community detection, we construct an undirected version of the graph and apply the Leiden algorithm (Traag, et al., 2019) with resolution 0.05, following the strategy in (Huong & Koch, 2025). The baseline partition yields 33 clusters with sizes reported in Figure 1.

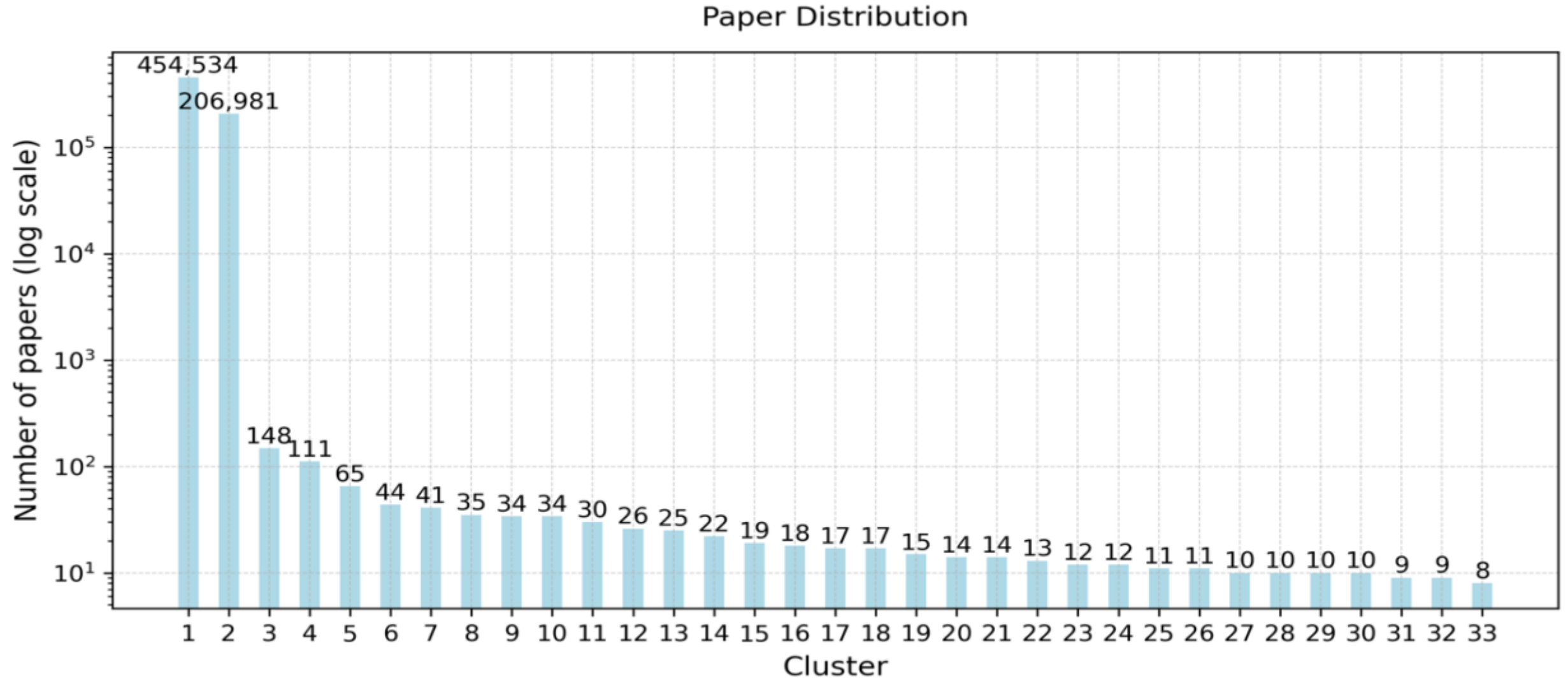

*Figure 1: Node distribution of the baseline solution*

To evaluate cluster quality, we compute *homogeneity*, defined as the fraction of cluster members that belong to the cluster's dominant subject category. A homogeneity of 100% means that all items in the cluster share the same label. As observed in (Huong & Koch, 2025), the two dominant clusters correspond closely to the two subject categories:

- Mathematics homogeneity ≈ 98%
- OR&MS homogeneity ≈ 86%

However, 31 additional clusters remain, with sizes ranging from 148 nodes to 8 nodes. Many of these smaller clusters exhibit nearly 100% subject homogeneity. The link distribution matrix among clusters in Figure 2 shows that most external edges from these small clusters connect primarily to the two largest clusters. This pattern suggests that their structural isolation is more likely due to missing citation connectivity rather than the presence of distinct intellectual domains.

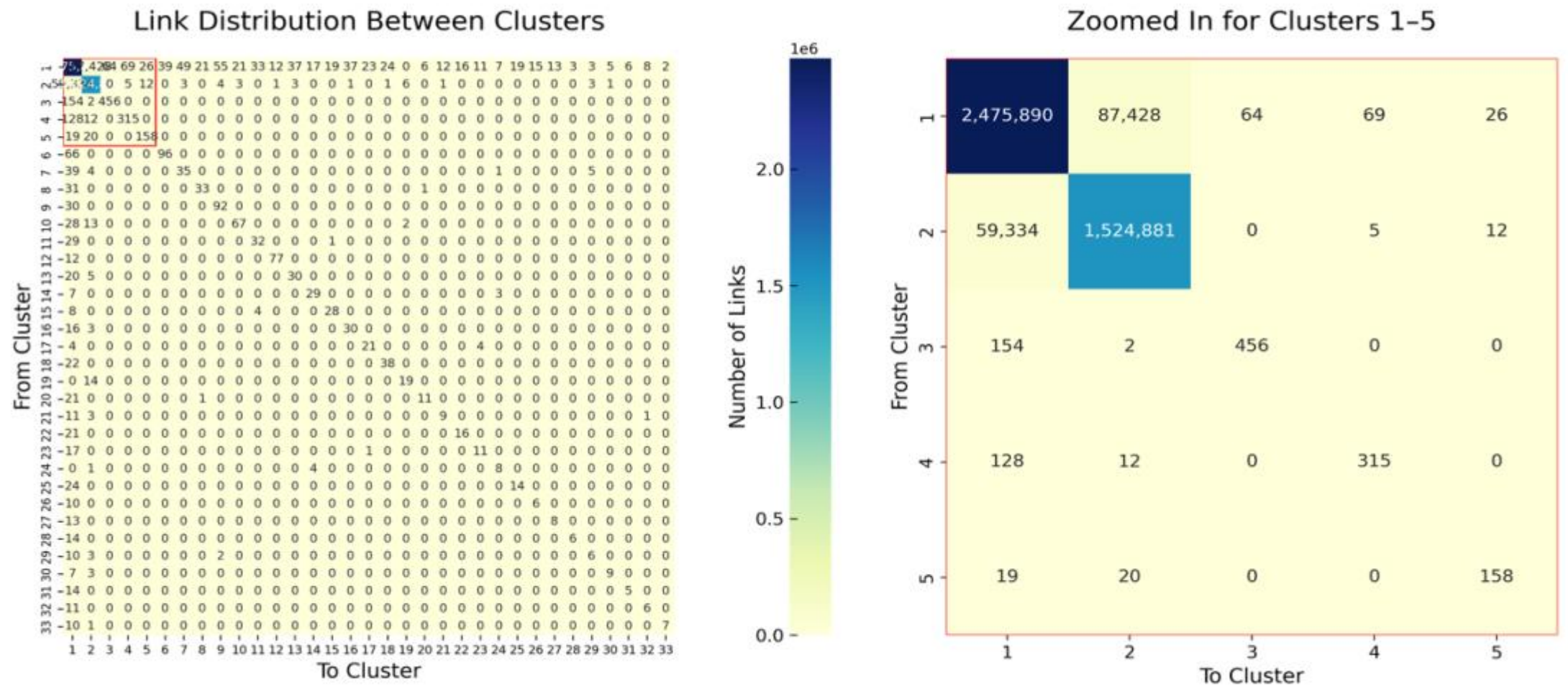

*Figure 2: Link distribution of the baseline solution*

To investigate this, we track the change from the number of references of articles in these small clusters (available in WoS metadata) to those retained in the extracted graph. Across 15,559 references, 10,502 (≈67.5%) were within WoS covergage. Limiting to the 2000–2024 period reduces this to 2,704 references (≈17.4%), and then 2,665 references (≈17.1%) appear as outgoing edges in the extracted graph, giving an overall retention of 17.13% (Figure 3). These results indicate that missing links in small clusters largely arise from limitations in database coverage, time-window truncation, and extraction artifacts.

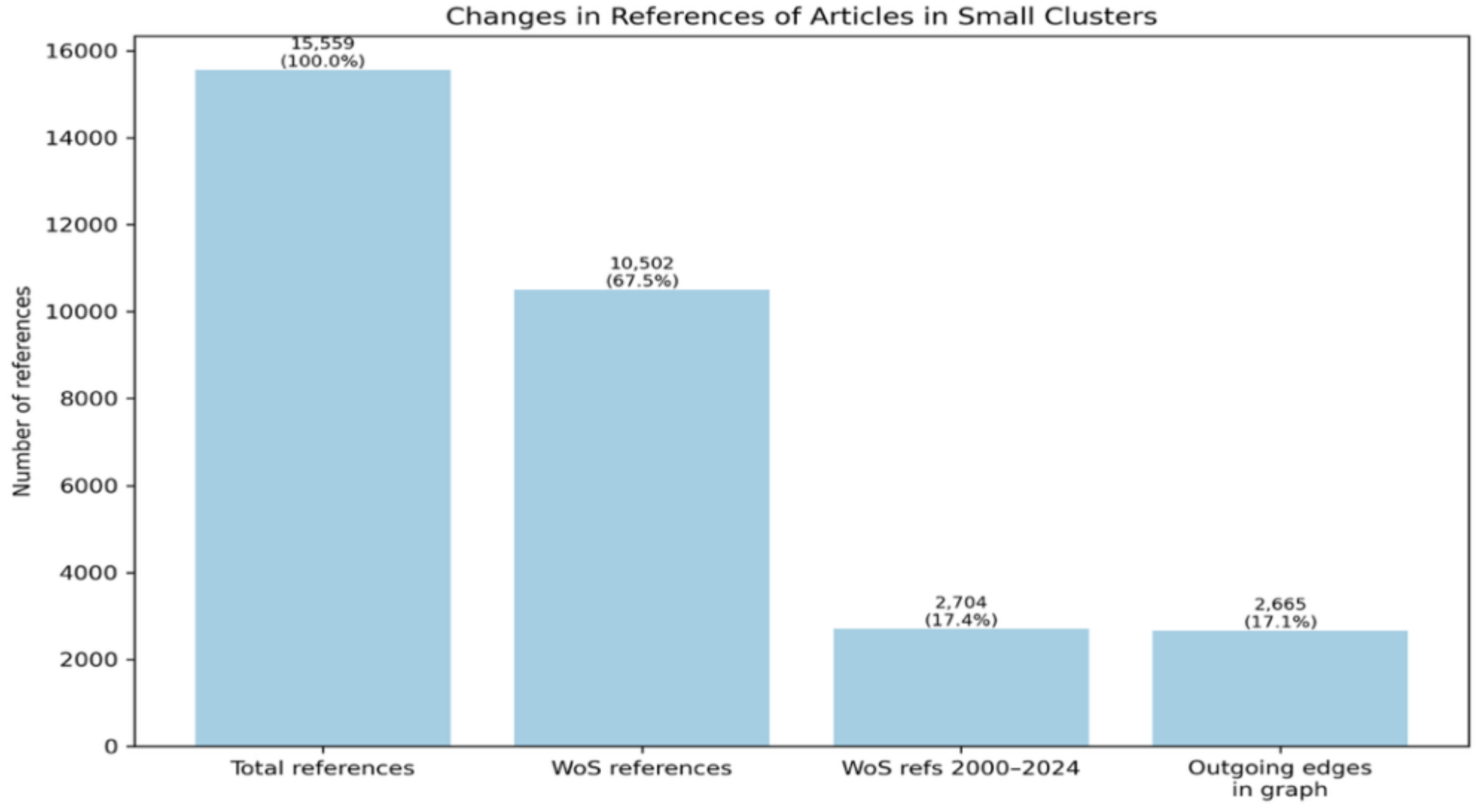

*Figure 3: Changes in references of publications in small clusters:*
*Coverage, Time Window, and Extraction Loss*

As an illustrative manual check, the article "*Some Remarks on Minimal Asymptotic Bases of Order Three*" (WOS:000546183200004, Cluster 4), published in Bull. Aust. Math. Soc. in 2020 by Ling Dengrong and Tang Min, has 9 references: 2 not in WoS coverage, 4 outside the years covered in our dataset (2000–2024), and 4 retained in the subgraph (all four in the Mathematics cluster).

Given these patterns of missing connectivity, publications in these small clusters are therefore the candidates for semantic repair.

## 3. Methods

### *3.1 Semantic Augmentation*

To introduce a complementary signal of relatedness, we compute semantic embeddings from titles and abstracts available in the WoS metadata. For each publication, title and abstract are concatenated and encoded using the large language model *mxbai-embed-large* (MixedbreadAI, 2025), producing a 1024-dimensional vector. We then use cosine similarity as the measure of semantic proximity. The interested reader is refered to (Kunt, et al., 2025) for a detailed discussions on embedding WoS data using different LLMs.

Importantly, we do not compute a fully dense textual graph. Instead, semantic information is selectively incorporated in two complementary steps.

**(S1) Repair of Small Components**

For nodes belonging to small clusters in the baseline graph, we construct semantic k-nearest-neighbor (kNN) edges. Similarity search is restricted to these nodes, i.e., it focuses on structurally vulnerable regions. For each selected publication, the top-k semantically closest neighbors are identified and added as edges. We experiment with k values of 10, 100, and 1000 and obtain three textual graphs with 854 nodes and 5815, 72762, and 834773 edges, respectively.

**(S2) Semantic Weighting of Citation Edges**

For existing citation edges, we compute the cosine similarity between the textual representations of the citing and cited publications. Each citation edge is then weighted according to this semantic similarity, thereby incorporating topical relatedness into the graph structure. As a result, citations between semantically similar papers receive higher weights, while semantically weak citations receive lower influence in the community detection process. The idea of semantic weighting is also explored in another of our works, addressing a broader scope within the research project "Fully Algorithmic Librarian" (FANs, 2026).

As a glimpse view of the weighting scheme, Figure 4 presents the distribution of the resulting edge weights. The values range from approximately 0.3 to 1.0 and exhibit a left-skewed distribution with a clear peak around 0.8, indicating that most citation pairs have relatively high semantic similarity. A small but noticeable fraction of edges receive substantially lower weights (0.3–0.6). This heterogeneity in edge weights motivates the use of semantic weighting rather than relying solely on the raw citation network to better reflect citation relations.

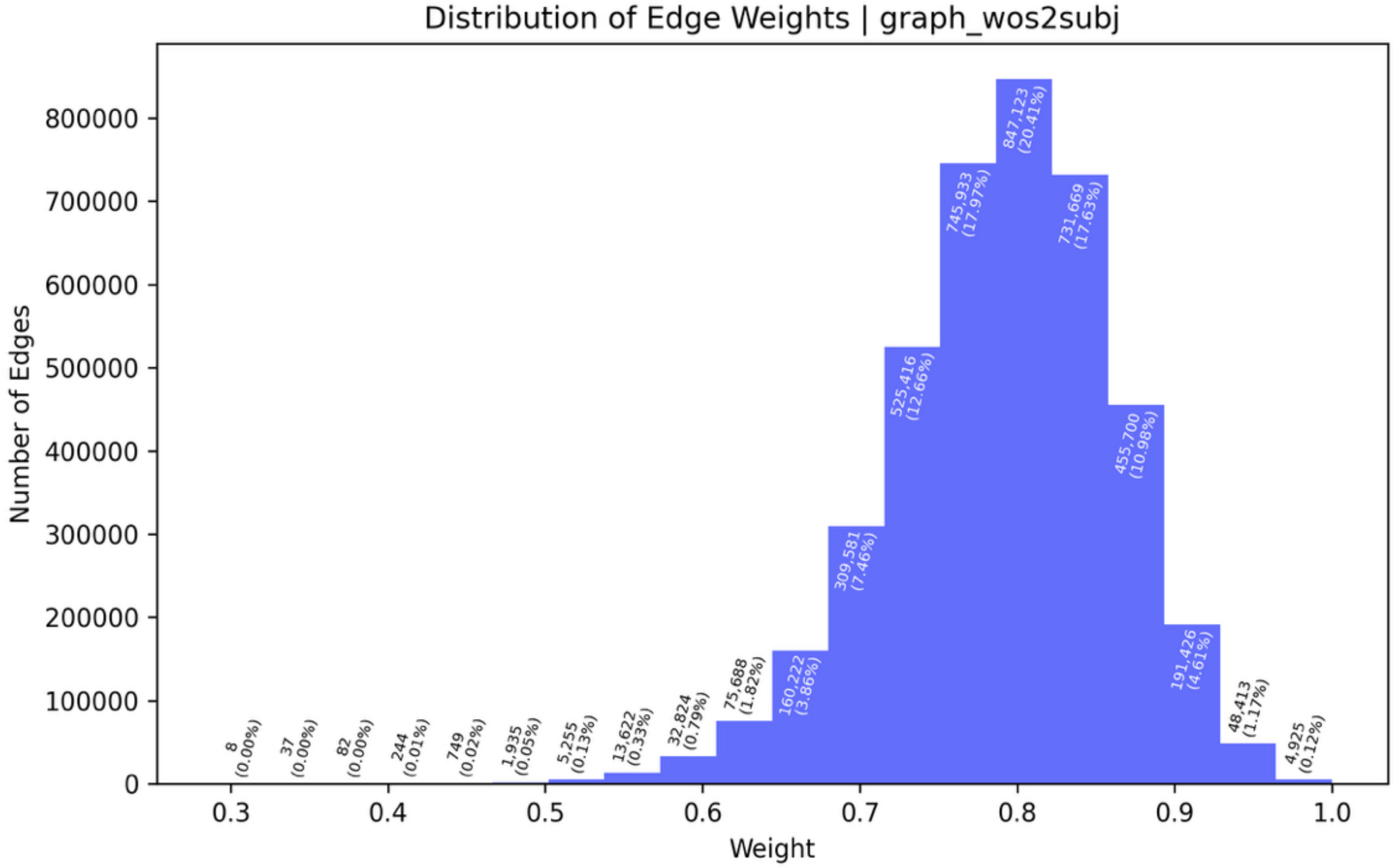

*Figure 4: Weight distribution of citations*

The textual edges obtained from (S1) are then merged with the citation edges from (S2), with an optional weighting parameter $\alpha \in [0,1]$, to form an augmented graph with edge weights

$$w_{ij} = \alpha\, w_{ij}^{textual} + (1-\alpha)\, w_{ij}^{citing}.$$

As a primary experiment, we set $\alpha = 0.5$ to obtain a balanced contribution from textual similarity and citation relationships. In practice, smaller values of $\alpha$ can also be used to reduce the influence of the artificially introduced semantic edges while preserving the dominant role of citation-based connections.

Finally, we construct three augmented graphs, denoted as G10nn, G100nn, and G1000nn, corresponding to the values $k$ (10, 100, and 1000). While the number of nodes remains the same as in the original graph G, the number of edges increases with larger $k$. There are 1,376, 2,266, and 2,526 edges appear in both the citing and textual graphs for $k = 10$, $k = 100$, and $k = 1000$, respectively, corresponding to 23.66%, 3.11%, and 0.30% of the textual edges.

*3.2 Community Detection and Comparison*

Leiden clustering is re-applied to the augmented graphs (with/without edge weights), and the results are compared with the baseline obtained by applying Leiden clustering to the original citation graph G. The comparison is based on several criteria: the number of clusters, the cluster size distribution, disciplinary homogeneity, and the stability of the dominant partitions.

In addition, we include a comparison with an embedding-only clustering approach by applying K-means (Jain, 2010) with a preset number of clusters as 2 to the 1024-dimensional embeddings.

## 4. Numerical Experiments

This section presents the empirical evaluation of the proposed semantic integration approach. We first examine how semantic augmentation reduces fragmentation in the citation network by progressively introducing semantic nearest-neighbor edges. We then analyze the effect of weighting citation edges by semantic similarity. Finally, we compare the hybrid citation-semantic graph approach with embedding-only clustering using K-means.

### *4.1 Fragmentation Reduction*

This subsection evaluates how semantic augmentation reduces fragmentation relative to the baseline citation graph G. We compare the clustering results obtained from the augmented graphs G10nn, G100, and G1000nn with the baseline Leiden partition. For comparability with the baseline setting, the Leiden algorithm is applied without incorporating cosine similarity as edge weights during clustering. Recall that with the baseline graph G, Leiden clustering identifies 33 clusters, although the structure is dominated by two large communities: Cluster 1 (454,534 papers) and Cluster 2 (206,981 papers).

**G10nn.** After semantic augmentation with k = 10 nearest neighbors, the resulting graph G10nn contains 4,155,291 edges, compared with 4,150,852 edges in the baseline graph. Despite the relatively small increase in edge count, Leiden clustering now produces only 9 clusters, indicating a substantial reduction in fragmentation. The two dominant clusters remain stable, containing 451,647 and 210,490 papers, respectively. The remaining clusters are small, ranging from 78 to 9 papers, indicating that most previously micro-components have been absorbed into the dominant communities. Importantly, disciplinary composition remains largely unchanged. The Mathematics cluster retains a homogeneity above 98%, while the OR&MS cluster remains around 85%.

**G100nn.** Increasing semantic connectivity to k = 100 produces the graph G100nn with 4,221,348 edges. Leiden clustering further reduces the number of clusters to 6. Again, the two dominant communities contain the vast majority of publications (455,135 and 207,120 papers), while the remaining clusters are small (76–11nodes). Disciplinary homogeneity remains stable, with Mathematics at roughly 98% homogeneity and OR&MS at approximately 86%.

**G1000nn.** For the densest augmentation (k = 1000), the graph G1000nn contains 4,982,625 edges, substantially strengthening connectivity across the network. In this configuration, Leiden identifies only two clusters, corresponding directly to the two disciplinary communities observed in the baseline graph. Cluster 1 contains 467,476 papers and is dominated by Mathematics with 97.1% homogeneity, while Cluster 2 contains 194,893 papers and is dominated by OR&MS with 89.4% homogeneity. The disappearance of all small clusters indicates that semantic augmentation effectively reconnects fragmented components while preserving the main disciplinary boundary.

Overall, increasing semantic connectivity progressively reduces fragmentation, from 33 clusters to 2 clusters. Across all augmentation levels, disciplinary homogeneity remains high (Mathematics >97%, OR&MS >85%), suggesting that semantic augmentation repairs structural gaps in the citation network without collapsing disciplinary distinctions.

*4.2 Effects of Semantic Weighting*

Introducing semantic weights slightly refines the clustering structure. Applying Leiden clustering to the weighted version of G1000nn yields four clusters, rather than the two observed in the unweighted version. Publications in the 31 small clusters from the baseline solution are absorbed into the two dominant clusters, consistent with the unweighted graph. Incorporating textual similarity, however, reveals two new small clusters (sizes 11 and 7) derived from Cluster 1 of the unweighted solution (the mathematics community). Among these nodes, 39 of 182 references (~22%) appear as outgoing edges in the citation graph, and they have no connections in the textual similarity graph. This suggests that these nodes could benefit from further augmentation.

These results are summarized in Table 1, which compares the cluster composition of the unweighted and weighted G1000nn graphs.

Table 1: Effect of semantic weighting on G1000nn cluster structure

| **Weighted G1000nn** | | | | **Unweighted G1000nn** | | | |
|---|---|---|---|---|---|---|---|
| Cluster | Mathematics | OR&MS | both | Cluster | Mathematics | OR&MS | both |
| 1 | 454,362 | 14,225 | 115 | 1 | 453,918 | 13,444 | 114 |
| 2 | 20,137 | 173,456 | 56 | 2 | 20,599 | 174,237 | 57 |
| 3 | 11 | 0 | 0 | – | – | – | – |
| 4 | 7 | 0 | 0 | – | – | – | – |

*4.3 Comparison with Embedding-Only Clustering*

To assess the value of incorporating citation structure, we compare the hybrid graph approach with embedding-only clustering. Specifically, K-means clustering with a preset number of clusters of 2 is applied directly to the 1024-dimensional text embeddings. Disciplinary homogeneity is comparable to the hybrid approach: the Mathematics cluster reaches 98.3%, while the OR&MS cluster reaches 84.8%. However, the distribution of misclassified papers differs quite significantly, particularly in the off-diagonal elements of the confusion matrix, see Table 2.

Table 2: Cluster composition of Kmeans compared with the hybrid approach on G1000nn

| **K-means** | | | | **Unweighted G1000nn** | | | |
|---|---|---|---|---|---|---|---|
| Cluster | Mathematics | OR&MS | both | Cluster | Mathematics | OR&MS | both |
| 1 | 442,287 | 7,692 | 90 | 1 | 453,918 | 13,444 | 114 |
| 2 | 32,230 | 179,989 | 81 | 2 | 20,599 | 174,237 | 57 |

Compared with the hybrid graph approach, K-means places a larger number of Mathematics papers into the OR&MS cluster. More importantly, embedding-only clustering has several limitations:

- The number of clusters must be fixed in advance;
- Citation topology is ignored;
- Multi-scale organization cannot be represented.

By contrast, the hybrid approach combines citation connectivity with semantic proximity, allowing the clustering algorithm to adaptively determine community structure. As shown in (Huong & Koch, 2025), Leiden clustering can reveal both coarse disciplinary partitions and smaller substructures depending on graph connectivity and resolution.

Taken together, these experiments show that semantic integration substantially improves the structural coherence of the citation network. Adding a limited number of semantic nearest-neighbor edges progressively reduces fragmentation while preserving the dominant disciplinary boundary between Mathematics and OR&MS. Introducing semantic weighting reveals new small groups that could benefit from further augmentation. Finally, comparison with embedding-only clustering shows that the hybrid approach achieves similar disciplinary homogeneity while retaining the structural richness of citation-based networks. Table 3 summarizes the clustering outcomes across all experimental settings.

Table 3: Summary of clustering outcomes for the baseline citation graph, semantically augmented graphs, and Kmeans

| **Graph / Method** | **Edges** | **Clusters** | **Cluster 1's Size** | **Cluster 2's Size** | **Mathematics Homogeneity** | **OR&MS Homogeneity** |
|---|---|---|---|---|---|---|
| G (baseline, unweighted) | 4,150,852 | 33 | 454,534 | 206,981 | 98.0% | 86.4% |
| G10nn (unweighted) | 4,155,291 | 9 | 451,647 | 210,490 | 98.1% | 85.2% |
| G100nn (unweighted) | 4,221,348 | 6 | 455,135 | 207,120 | 98.0% | 86.3% |
| G1000nn (unweighted) | 4,982,625 | 2 | 467,476 | 194,893 | 97.1% | 89.4% |
| G1000nn (weighted) | 4,982,625 | 4 | 468,702 | 193,649 | 96.9% | 89.6% |
| K-means (embeddings) | – | 2 | 450,069 | 212,300 | 98.3% | 84.8% |

## 5. Computational Cost and Scalability

Constructing the full textual graph by computing pairwise similarities among all N nodes requires

$$O(N^2) \approx 438\text{B}$$

similarity computations. Such a quadratic cost is computationally expensive and becomes impractical for large graphs. Our proposed hybrid method significantly reduces this cost by avoiding the computation of full pairwise similarities.

**Repair mode (S1).** Similarity search is restricted to a small subset S of nodes corresponding to small disconnected components. For each of these nodes, we perform a k-nearest-neighbor search over the N nodes. Using standard nearest-neighbor search libraries, the complexity is

$$O(S \log N), \quad \text{where } S \ll N$$

(the $\log N$ factor reflects the cost of kNN searches). Even if using a brute-force similarity computation for S nodes, the cost of the repair mode would be $O(SN)$, which is linear w.r.t. the number of total nodes, significantly smaller than $O(N^2)$.

**Weighting mode (S2).** Semantic similarity is computed only for existing citation edges in the original graph. Therefore, the complexity is proportional to the number of citation edges:

$$O(E_{citing}) \approx 4.15\text{M}.$$

It is worth noting that the number of citation edges can be approximated from the number of nodes $N$ in the graph:

$$E_{citing} \approx RN$$

where $R$ is the average number of references per publication ($R = 14$ reported in Table 1 of (Chen, et al., 2023) for a citation graph of all WoS subjects). While $N$ can grow substantially – up to 60 million if the entire WoS dataset is considered – $R$ remains relatively stable due to consistent citation practices in science.

Overall, the proposed semantic augmentation method reduces the computational cost from quadratic $O(N^2)$ to at *most linear in the number of nodes*:

$$O(SN + E_{citing}) \approx O((S + R)N),$$

making the approach highly scalable to large datasets.

## 6. Conclusion

This study demonstrates that LLM-based semantic similarity can compensate for structural fragmentation in large citation graphs without distorting disciplinary structure. Small disconnected components in citation networks often reflect missing connectivity rather than novel topics. Targeted semantic augmentation reconnects these regions while preserving established field boundaries. Compared to embedding-only clustering, the hybrid framework maintains citation-based interpretability and multi-scale structures. Rather than replacing citation networks, semantic representations strengthen them where they are weakest. For broader bibliometric research, this provides a scalable and methodologically transparent approach to improving the robustness of citation-based indicators at WoS scale.

**Acknowledgements.** This work is co-funded by the European Union (European Regional Development Fund EFRE, Fund No. STIIV-001 and STIIV-004) and supported by the German Competence Network for Bibliometrics (Grant No. 16WIK2101A).

**Data and Code Availability.** The bibliographic data used in this study were obtained from the Web of Science through the subscription provided by the German Competence Network for Bibliometrics and therefore cannot be publicly redistributed. The analyses rely on standard implementations in widely used open-source libraries: k-means clustering and k-nearest neighbors were implemented using scikit-learn (https://scikit-learn.org/stable/), while Leiden community detection was implemented using igraph (https://igraph.org/python/) and leidenalg (https://github.com/vtraag/leidenalg).

**Competing interests.** The authors declare that they have no competing interests. All authors have read and approved the manuscript.